\documentstyle[12pt]{ioplppt}          
\begin{document}
\title{Hamilton-Jacobi Solutions for Strongly-Coupled Gravity and Matter}
\author{D.S. Salopek}
\address{Department of Physics \& Astronomy\\
6224 Agricultural Road, University of British Columbia\\
Vancouver, Canada V6T 1Z1}
\begin{abstract}
A Green's function method is developed for solving
strongly-coupled gravity and matter in the semiclassical limit.
In the strong-coupling limit, one assumes that Newton's constant
approaches infinity, $G \rightarrow \infty$. As a result,
one may neglect second order spatial gradients, and each spatial point
evolves like an homogeneous universe. After constructing the
Green's function solution to the Hamiltonian
constraint, the momentum constraint is solved using
functional methods in conjunction with the superposition principle
for Hamilton-Jacobi theory. Exact and approximate solutions are given
for a dust field or a scalar field interacting with gravity.
\end{abstract}

Strong-coupling expansions have proven to been useful in
elucidating gauge theories of modern particle physics
(see, {\it e.g.}, Creutz 1983 and Polyakov 1987). A general
class of Hamilton-Jacobi solutions for strongly-coupled
gravity and matter will be presented here. It is hoped that
these solutions will help clarify quantum features of gravity
that are necessary in formulating the inflationary scenario
(Hartle \& Hawking 1983 and Hartle 1997).

In the context of classical gravity, the strong-coupling expansion is
very similar although not exactly the same as a long-wavelength
expansion. In their analysis of singularities in the early Universe,
Lifshitz and Khalatnikov (1964) expanded the 3-metric in a Taylor series
in what was basically a strong-coupling expansion
(see also Belinski \etal 1970 and Landau and Lifshitz 1975).
However, their program was incomplete in that they did not
give an explicit solution of the momentum constraint equation
which is just the $G^0_i$ Einstein equation. Pilati (1982),
Teitelboim (1982) and Henneaux {\it et al} (1982)
took some initial steps in formulating
a quantum  treatment of strongly-coupled gravity although many
difficulties still remain (see also Husain 1988).
They pointed out that this system
is much simpler than the case for full gravity: the Hamiltonian
was similar to that of an homogeneous minisuperspace model. Solutions
were given which were a product over the points on a lattice.
Once again, they did not attempt to solve the momentum constraint.
In the present work, I hope to improve upon this situation
by showing how to simultaneously solve both the energy and momentum constraints
in the semiclassical limit. Special cases of the semiclassical
limit were given before (Salopek 1991), but now general solutions will be
given.

A long-wavelength expansion
has been considered by numerous authors, including Tomita (1975),
Salopek and  Stewart (1992), Deruelle and Langlois (1995) and
Veneziano (1997). However,
by using Hamilton-Jacobi (HJ) methods, one can elegantly
solve the momentum constraint. In fact, Parry \etal (1994)
used a spatial gradient expansion which is manifestly
invariant under spatial coordinate transformations to
give a HJ solution for gravity and matter. They described
a systematic method of obtaining terms of arbitrarily
high order. Soda \etal (1995), Chiba (1995) and Darian (1997)
have extended this method to
encompass Brans-Dicke gravity, n-dimensional gravity, and
gravity interacting with electromagnetism, respectively.

In a previous paper, Salopek (1991) had shown that one could construct
solutions of the energy constraint and the momentum constraint for
strongly-coupled gravity by assuming that
the generating functional was of the simple form:
\begin{equation}
{\cal S} = -2 \int d^3 x \; \gamma^{1/2} \; H[\phi(x), \chi(x)] \, ,
\label{simple}
\end{equation}
where $\phi(x)$ and $\chi(x)$ denote the matter fields.
If the Hubble function $H$ were a constant, then this functional
would correspond to the volume of any given 3-geometry.
For many astrophysical applications, the ansatz of eq.(\ref{simple})
is sufficient.
Such a formalism provides an elegant description of the
long-wavelength evolution of fluctuations arising from inflation.
For example, it was used to construct inflation models that yielded
non-Gaussian fluctuations (Salopek 1992a). Such models are still
of observational and theoretical interest (Moscardini \etal 1993).
However, for future extensions
of the HJ approach,  it is of interest to construct a more general class of
solutions. Some steps were taken in that direction by Salopek (1991)
who presented
a complete solution of the energy constraint of the long-wavelength
problem. In addition, by a suitable choice
of the spatial coordinates, one could solve the momentum constraint.
In the present work, powerful geometric methods will be utilized
to give a more elegant solution of the momentum constraint.

Over the past few years, an ever increasing number of tools
have been developed in order to solve the functional
Hamilton-Jacobi equation. They include:

\noindent
{\it (1) The Spatial Gradient Expansion.}
For centuries, Taylor series solutions have been constructed for
linear equations. Series expansion techniques can also
be applied to nonlinear differential equations including the
HJ equation for general relativity (Parry \etal 1994, Salopek 1997).
In this way, it is possible to decompose semiclassical superspace
into a sum of minisuperspaces.

\noindent
{\it (2)  The Superposition Principle.}
Since the equations for quantum mechanics are linear, the superposition
principle plays an important role. By employing the stationary
phase approximation, one can also enunciate a Superposition Principle
for Hamilton-Jacobi theory (see, {\it e.g.}, Salopek 1997).
Although it is not linear, it proves very effective in constructing
complicated solutions from more elementary ones (Landau \& Lifshitz 1960).
It will also prove useful in solving inhomogeneous gravitational problems such
as the one presented in this paper.

The Lagrangian and Hamiltonian formulations for strongly-coupled gravity
and matter are given in section 1. The semiclassical solution method may be
simply explained using some rudimentary
ideas from quantum mechanics. A complete solution of the energy constraint
for a dust field interacting with gravity is described in section 2.
This solution is related to the well-known Kasner metric.
The momentum constraint is solved using a superposition over the complete
set of solutions. It is also shown how to construct the set of
{\it constant functionals}. Explicit solutions are constructed for a
single dust field in section 3
and for two dust fields in section 4. In sections 5, 6 and 7,
the entire analysis is repeated for a scalar field interacting with
gravity. Conclusions are given in section 8.

\section{Strongly-Coupled Equations for Gravity and Matter}

The action for Einstein gravity interacting with a dust field,
$\chi$, and a scalar field, $\phi$, may be written as follows:
\numparts
\begin{equation}
{\cal I} = \int d^4x \, \sqrt{ -g} \; \left \{
{1 \over 2 \kappa } {}^{(4)}R
- {1 \over 2 \kappa } g^{\mu \nu} \phi_{,\mu} \phi_{,\nu} - \kappa V(\phi)
- {n \over 2}
( g^{\mu \nu} \chi_{,\mu} \chi_{,\nu} + \kappa^2 ) \right \} \, ,
\label{action}
\end{equation}
where
\begin{equation}
\kappa \equiv  8 \pi G = 8 \pi/ m_P^2 \,
\end{equation}
\endnumparts
is the gravitational coupling constant.
$n \equiv n(t,x)$ is a Lagrange multiplier which
ensures that the square of the four-velocity, $U^{\mu}$, is minus one:
\begin{equation}
U^\mu \, U_\mu = - 1 \, , \quad U^\mu =
- g^{\mu \nu} \, \chi_{,\nu} / \kappa\, .
\end{equation}
The above form can be obtained from the usual one by appropriate
scalings of the matter fields and their coupling constants.
Factors of $\kappa$ appear
in key places of eq.(\ref{action}) in order to obtain the desired
energy constraint:
\numparts
\begin{eqnarray}
{\cal H}(x) &&= \kappa \; \gamma^{-1/2} \,
\left ( 2 \gamma_{ac} \gamma_{bd} - \gamma_{ab} \gamma_{cd} \right )
\pi^{ab} \pi^{cd}
+ \kappa { \gamma^{-1/2} \over 2 }  \left ( \pi^{\phi} \right )^2
+ \nonumber \\
&& \kappa \; \pi^\chi \sqrt{ 1 + \chi^{|a} \chi_{|a}/ \kappa^2 }
 -{\gamma^{1/2}  \over 2 \kappa} \;   R +
{\gamma^{1/2} \over 2 \kappa} \phi^{|a} \phi_{|a} +
\kappa  \; \gamma^{1/2} \, V(\phi) = 0\; .
\label{ec}
\end{eqnarray}
No factors of $\kappa$ appear in the momentum constraint:
\begin{equation}
{\cal H}_{i}(x)=-2\left(\gamma_{ik} \pi^{kj} \right)_{,j} +
\pi^{kl} \gamma_{kl,i} + \pi^\phi \phi_{,i} +
\pi^\chi \chi_{,i} = 0 \, .
\label{mc}
\end{equation}
\endnumparts
In a classical Hamilton-Jacobi formulation of general relativity, one defines
the generating functional,
\numparts
\begin{equation}
{\cal S} \equiv {\cal S}[\gamma_{ab}(x), \phi(x), \chi(x)],
\end{equation}
by assigning a real number to each field configuration $[\phi(x),\chi(x)]$
on a space-like hypersurface with 3-geometry given by
$\gamma_{ab}(x)$. In a semiclassical context, one allows for the possibility
that ${\cal S}$ may be complex.
The Hamilton-Jacobi equations are obtained
from the constraints (\ref{ec}-b) by replacing the
momenta with functional derivatives of ${\cal S}$:
\begin{equation}
\pi^{\chi}(x) = { \delta {\cal S} \over \delta \chi(x) }, \quad
\pi^{\phi}(x) = { \delta {\cal S} \over \delta \phi(x) }, \quad
\pi^{ab}(x) = { \delta {\cal S} \over \delta \gamma_{ab}(x) } \, .
\end{equation}
\endnumparts

In the limit of large gravitational coupling,
$\kappa \rightarrow \infty$, the generating functional ${\cal S}^{(s)}$
for the strongly-coupled system evolves according to the following equations:
\numparts
\begin{eqnarray}
{\cal H}^{(s)}(x)/ \kappa
= && \;
\gamma^{-1/2}\,
\left ( 2 \gamma_{ac} \gamma_{bd} - \gamma_{ab} \gamma_{cd} \right )
{\delta {\cal S}^{(s)} \over \delta \gamma_{ab}}
{\delta {\cal S}^{(s)} \over \delta \gamma_{cd}}
 +  \; {\delta {\cal S}^{(s)} \over \delta \chi}  \nonumber \\
&& + {\gamma^{-1/2} \over 2 }
\left ( {\delta {\cal S}^{(s)} \over \delta \phi} \right )^2
+    \; \gamma^{1/2} \, V(\phi)  \; = 0 \; ,
\label{ecs}
\end{eqnarray}
\begin{equation}
{\cal H}^{(s)}_{i}(x)=-2\left(\gamma_{ik} \,
{\delta {\cal S}^{(s)} \over \delta \gamma_{kj}} \right)_{,j} +
{\delta {\cal S}^{(s)} \over \delta \gamma_{kl}} \gamma_{kl,i}
+ { \delta {\cal S}^{(s)} \over \delta \phi } \, \phi_{,i} +
{ \delta {\cal S}^{(s)} \over \delta \chi } \, \chi_{,i} = 0 \, .
\label{mcs}
\end{equation}
\endnumparts
The energy constraint is ultra-local in the sense that different
spatial points are not coupled to each other. As a result,
the Poisson bracket of ${\cal H}^{(s)}(x)$ with
${\cal H}^{(s)}(y)$ vanishes, and consistency  of the Hamiltonian
constraint at different spatial points is assured.
Consistency or `integrability' of the Hamiltonian constraint for full general
relativity is more complicated. It is related to the freedom in choosing
an arbitrary time foliation: see Parry \etal (1994)
as well as Salopek (1995). However, the momentum
constraint does couple the spatial points together because spatial
derivatives appear.  It is this cross-coupling of different
spatial points that makes the strongly-coupled system non-trivial.

(In order to simplify the notation,  ${\cal S}$ will be used to denote
${\cal S}^{(s)}$ for the remainder of this paper.)

\subsection{Analogy from Elementary Quantum Mechanics}

The general solution for the strongly-coupled system
may appear quite complicated
so it is instructive to consider a well-known example from elementary
quantum mechanics which illustrates the essential features.

\subsection{Rotationally Symmetric Solutions to the Schrodinger Equation for
a Free Particle}

The two-dimensional Schrodinger equation for a free particle with zero
angular momentum is given by:
\numparts
\begin{equation}
i { \partial \psi \over \partial t} = - { 1 \over 2 m} \left (
{\partial^2 \psi \over \partial x^2} + {\partial^2 \psi \over \partial y^2}
\right ) \, , \label{schrod}
\end{equation}
\begin{equation}
L_z \; \psi = i \left ( y { \partial \over \partial x}
- x { \partial \over \partial y} \right ) \psi = 0 \, .
\label{symmetry}
\end{equation}
\endnumparts
In solving these two equations, one would ordinarily invoke polar coordinates,
$(r, \theta)$, and then write the solution as $\psi \equiv \psi(r)$.
Instead, I will utilize a circuitous method which is of
interest because the same technique may be generalized to the case of
strongly-coupled gravity.

One notes immediately that the first equation (\ref{schrod})
can be solved generally without any reference to the second by using
a superposition of plane waves:
\begin{equation}
\Psi = \int d^2k \; f(\vec k) \;
e^{i ( \vec k \cdot \vec x - \omega t ) } \,
\quad w = (k_1^2 + k_2^2)/ (2m),
\end{equation}
where $f(\vec k)$ is an arbitrary function of the
wave-vector $\vec k$. Please note that a plane wave
$e^{i ( \vec k \cdot \vec x - \omega t ) }$
is {\it not} a solution of the symmetry condition,
eq.(\ref{symmetry}). However, by suitably restricting the function
$f$ so that it is function only of the magnitude of
$k = \sqrt{k_1^2 + k_2^2}$,
\begin{equation}
f(\vec k) \equiv f(k) \,
\end{equation}
one can indeed satisfy the symmetry condition.

In a semiclassical gravitational context,
the energy constraint (\ref{ecs}) will
be analogous  to the Schrodinger equation (\ref{schrod}) and the momentum
constraint (\ref{mcs}) will be analogous to the symmetry condition
(\ref{symmetry}). In general relativity, it is not yet known
how to explicitly solve the
momentum constraint such that one obtains a set of reduced
variables ({\it physical variables}) analogous to solving the $L_z$
constraint and then deducing that $\psi$ is solely a function of $r$.
Instead, one solves the strongly-coupled system by
first obtaining a general class of solutions
for the energy constraint. By considering a suitable superposition
over these solutions, one then constructs a solution to the momentum
constraint.

Shortly after Dirac (1958) gave general relativity its
Hamiltonian form, Higgs (1958) pointed out that the momentum constraint
of general relativity implied that the wavefunctional was
invariant under spatial coordinate transformations.
Although, the Hamilton-Jacobi equation was written explicitly by Peres (1962),
only quite recently (Salopek and Stewart 1992, Parry \etal 1994)
has the spatial coordinate invariance been
exploited to construct explicit solutions of the full HJ equation.
They utilized a spatial gradient expansion.

\section{Strongly-Coupled Solutions: Dust Interacting with Gravity}

In the strongly-coupled limit, the energy constraint for gravity and
a dust field is:
\begin{equation}
{\cal H}^{(s)}(x)/ \kappa =
\gamma^{-1/2} \, \left (
2 \gamma_{ac} \gamma_{bd} - \gamma_{ab} \gamma_{cd}
\right )
{\delta {\cal S} \over \delta \gamma_{ab}}
{\delta {\cal S} \over \delta \gamma_{cd}} +
{\delta {\cal S} \over \delta \chi}  = 0 \, .
\label{hjsdg}
\end{equation}

\subsection{Green's Function Solution for the Strongly-Coupled HJ Equation}

A complete solution, ${\cal G}$, to the above equation is:
\numparts
\begin{eqnarray}
{\cal G}[ \gamma_{ab}(x), \chi(x) | && \gamma^{(0)}_{ab}(x), \chi_{(0)}(x) ]
= {4 \over 3 } \int d^3 x \;
{1 \over \left ( \chi(x) - \chi_{(0)}(x) \right ) } \;  \nonumber \\
&& \left [ 2 \gamma^{1/4} \; \gamma_{(0)}^{1/4} \;
{\rm cosh} ( \sqrt{3 \over 8} z )
- \gamma^{1/2} - \gamma_{(0)}^{1/2} \right ] \, ,
\label{greenf}
\end{eqnarray}
where
\begin{equation}
z = \; {1 \over 2} \; \sqrt{ {\rm Tr}
\left \{
\ln \left( [h] [h_{(0)}^{-1}] \right )
\ln \left( [h] [h_{(0)}^{-1}] \right )
\right \} } \, ,
\label{defz}
\end{equation}
and $[h]$ and $[h^{(0)}]$ are matrices with components given by
\begin{equation}
[h]_{ab} =\gamma^{-1/3} \gamma_{ab} \, , \quad
[h^{(0)}]_{ab} =\gamma_{(0)}^{-1/3} \gamma^{(0)}_{ab} \, .
\end{equation}
\endnumparts
This solution will be referred to as the {\it Green's function}.
It depends on 7 {\it parameter fields}, $\chi_{(0)}(x)$ and
$\gamma^{(0)}_{ab}(x)$, which may be interpreted as the
{\it initial fields} for the dust and 3-metric.
Although the variable $z$ had been defined in an earlier paper
(Salopek 1991),
the present form for the generating functional ${\cal G}$ is new. It is
superior to the previous formulation in that it
is symmetric upon interchange of the metric
$\gamma_{ab}(x)$ with the initial metric $\gamma^{(0)}_{ab}(x)$:
\begin{equation}
{\cal G}[\gamma_{ab}(x), \chi(x) |
\; \gamma^{(0)}_{ab}(x), \chi_{(0)}(x) ]  =
{\cal G}[\gamma^{(0)}_{ab}(x), \chi(x) | \; \gamma_{ab}(x), \chi_{(0)}(x)]
\, .
\end{equation}
One may verify by differentiation that eq.(\ref{greenf})
is indeed a solution to  eq.(\ref{hjsdg}) but in its present form
it does not satisfy the momentum constraint,
\begin{equation}
{\cal H}^{(s)}_{i}(x)=-2\left(\gamma_{ik}
{\delta {\cal S} \over \delta \gamma_{kj} } \right)_{,j} +
{\delta {\cal S} \over \delta \gamma_{kl} } \, \gamma_{kl,i} +
{\delta {\cal S} \over \delta \chi } \, \chi_{,i} = 0 \, , \label{mcdg}
\end{equation}
because the parameter fields are spatially dependent and hence they
carry momentum.

\subsubsection{Too Many Parameter Fields}

There is an additional problem with the Green's function
solution eq.(\ref{greenf}) as it stands: there is one too many parameter
fields. This subtle fact is not immediately apparent but it becomes painfully
clear after one explicitly tries to implement the superposition
principle in Section 2.2. One can resolve this problem by arbitrarily
setting one of the parameter fields to zero, say $\chi_{(0)}(x)=0$.
A brief explanation of this point now follows although it is
best to skip over the next two paragraphs upon first reading of this
paper.

For each spatial point, there were originally seven degrees of
freedom associated with $\chi(x)$ and the symmetric 3-metric $\gamma_{ab}(x)$.
Because of the symmetry between the initial fields,
$[\gamma^{(0)}_{ab}(x), \chi_{(0)}(x) ]$ and the original fields,
there is an equal number of
degrees of freedom associated with the former. I will now demonstrate
that the initial fields are not really independent. First note that
there is a relationship between
\begin{equation}
\pi^{\chi}_{(0)}(x) \equiv - { \delta {\cal G} \over \delta \chi_{(0)}(x) }
\quad {\rm and} \quad \pi_{(0)}^{ab}(x) \equiv
- {\delta {\cal G} \over \delta \gamma^{(0)}_{ab}(x) }  \, ,
\end{equation}
because the Hamiltonian constraint
is also valid when expressed in terms of the initial fields:
\begin{equation}
0= {\cal H}^{(s)}(x)/ \kappa =
\gamma_{(0)}^{-1/2} \, \left ( 2 \gamma^{(0)}_{ac} \gamma^{(0)}_{bd} -
\gamma^{(0)}_{ab} \gamma^{(0)}_{cd} \right ) \;
\pi^{ab}_{(0)} \, \pi^{cd}_{(0)} \;
+ \pi_{(0)}^{\chi} \, .
\label{hjdg0}
\end{equation}
If one were to apply the superposition principle in Section 2.2
by minimizing ${\cal G} + {\cal Q}$ with respect to $\gamma^{(0)}_{ab}(x)$
{\it and} $\chi_{(0)}(x)$, one would find that
\numparts
\begin{eqnarray}
0 = && {\delta {\cal G} \over \delta \gamma^{(0)}_{ab} }  +
{\delta {\cal Q} \over \delta \gamma^{(0)}_{ab} } \, , \\
0 = && {\delta {\cal G} \over \delta \chi_{(0)} }  +
{\delta {\cal Q} \over \delta \chi_{(0)} } \, .
\end{eqnarray}
\endnumparts
As a result, the momenta
\numparts
\begin{eqnarray}
\pi^{\chi}_{(0)}(x) =&& {\delta {\cal Q} \over \delta \chi_{(0)}(x) } \, ,  \\
\pi^{ab}_{(0)}(x) = &&  {\delta {\cal Q} \over \delta \gamma^{(0)}_{ab}(x) }
\, ,
\end{eqnarray}
\endnumparts
are functions of $\gamma^{(0)}_{ab}(x)$ and $\chi_{(0)}(x)$.
Hence eq.(\ref{hjdg0}) implicitly defines a relationship amongst the
initial fields $[\gamma^{(0)}_{ab}(x), \chi_{(0)}(x)]$: they are not
independent variables.

In order to avoid the above problem I will now arbitrarily
fix one of the parameter fields. In the present work, I will choose
$\chi_{(0)}(x) = 0$, while $\gamma^{(0)}_{ab}(x)$ will remain
a free variable. The subsequent analysis is indeed
consistent, although other choices are presumably possible.

\subsection{Superposition Principle for HJ Theory and Solution of the
Momentum Constraint}

Assuming $\chi_{(0)}(x)=0$ in the general solution to the Hamiltonian
constraint,  eq.(\ref{greenf}-c),
one can construct a solution ${\cal S}$ to the momentum constraint through
a superposition principle:
\numparts
\begin{equation}
{\cal S}[\gamma_{ab}(x), \chi(x)] =
{\cal G}[\gamma_{ab}(x), \chi(x)| \; \gamma^{(0)}_{ab}(x),  \chi_{(0)}(x)=0] +
{\cal Q}[\gamma^{(0)}_{ab}] \, , \label{cals}
\end{equation}
where $\gamma^{(0)}_{ab}(x)$ has been chosen to minimize
${\cal G} \; + \; {\cal Q}$,
\begin{equation}
0 = {\delta {\cal G} \over \delta \gamma^{(0)}_{ab} }  +
{\delta {\cal Q} \over \delta \gamma^{(0)}_{ab} } \, .
\label{minimize}
\end{equation}
Here ${\cal Q}$ is an arbitrary ``gauge-invariant'' functional of
the initial 3-metric, $\gamma^{(0)}_{ab}$:
\begin{equation}
0= -2\left(\gamma^{(0)}_{ik}
{\delta {\cal Q} \over \delta \gamma^{(0)}_{kj} } \right)_{,j} +
{\delta {\cal Q} \over \delta \gamma^{(0)}_{kl} }
\gamma^{(0)}_{kl,i} \, .
\end{equation}
Construction of solutions to the strongly-coupled problem through the
superposition principle eq.({\ref{cals}-c) will be known
as the {\it Semiclassical Green's Function Method}.

An elementary discussion of the Superposition Principle for
HJ theory was given by Salopek (1997). As will be demonstrated
in section 3.3, the functional
${\cal Q}[\gamma^{(0)}_{ab}]$  may be interpreted as the initial state
for ${\cal S}$ when $\chi(x) = 0$:
\begin{equation}
{\cal S}[\gamma_{ab}(x), \chi(x)=0 ] = {\cal Q}[\gamma_{ab}] \, .
\label{initialstate}
\end{equation}
\endnumparts
Since the Poisson bracket of ${\cal H}^{(s)}_i(x)$ with
${\cal H}^{(s)}(x)$ vanishes
(weakly), the momentum constraint is preserved upon evolution:
if ${\cal S}$ satisfies ${\cal H}^{(s)}_i(x)=0$ for $\chi(x) = 0$, it is
guaranteed to solve it for arbitrary $\chi(x)$.

\subsection{Classical Evolution}

Classical evolution is obtained from the minimization prescription
eq.(\ref{minimize}) by solving for $\gamma_{ab}(x)$ in terms of
$\gamma_{ab}^{(0)}(x)$ and $\chi(x)$:
\numparts
\begin{equation}
\left( { \gamma \over \gamma_{(0)}} \right )^{1/2} =
\left( 1- {1 \over 2} \chi \, \gamma_{(0)}^{-1/2} \pi_{(0)} \right )^2
- { 3 \over 2} \; \chi^2 \gamma^{-1}_{(0)}
\; \overline \pi_{(0)}^{ab} \overline \pi^{(0)}_{ab} \; , \label{dgcea}
\end{equation}
\begin{equation}
z = \sqrt{8 \over 3 } \; {\rm tanh}^{-1}
\left \{
\sqrt{3 \over 2} \, \chi \,
\left( \gamma^{-1}_{(0)} \,
\overline \pi_{(0)}^{ab} \;
\overline \pi^{(0)}_{ab} \right )^{1/2} \, \Bigg /  \,
\left( 1 - {1 \over 2} \; \chi \gamma_0^{-1/2} \pi_0 \right ) \right \}
\, , \label{dgceb}
\end{equation}
\begin{equation}
[h] = [h^{(0)}] \; {\rm exp}
\left [
{2z \; [\overline \pi_{(0)}] \; [\gamma^{(0)}] \over
\left( \overline \pi_{(0)}^{ab} \;
\overline \pi^{(0)}_{ab} \right )^{1/2} } \, \right ] \; . \label{dgcec}
\end{equation}
Matrix notation denoted by $[\quad ]$ has been used to simplify
the above expression. For e.g.,  $[\gamma^{(0)}]$ denotes the 3-metric
with components
$\gamma_{ab}^{(0)}$ and $[h]$ denotes $h_{ab}$, {\it etc}. In addition
$[\pi_{(0)}(x)]$ is a matrix whose components
are just the functional derivative of ${\cal Q}$ with respect
to $\gamma_{ab}^{(0)}(x)$,
\begin{equation}
[\pi_{(0)}(x)]^{ab} \equiv \pi^{ab}_{(0)}(x) =
- {\delta {\cal G} \over \delta \gamma^{(0)}_{ab}(x)} =
{ \delta {\cal Q} \over \delta \gamma^{(0)}_{ab}  } \, , \label{dgced}
\end{equation}
\endnumparts
where the last equality follows from eq.(\ref{minimize}).
$[\overline \pi_{(0)}]$ is its traceless counterpart where
$\pi_{(0)}$ denotes the trace:
\begin{equation}
\pi_{(0)} = \gamma_{ab}^{(0)} \, \pi^{ab}_{(0)} \, , \quad
\overline \pi^{ab}_{(0)} = \pi^{ab}_{(0)}
- {1 \over 3} \, \gamma_{(0)}^{ab} \, \pi_{(0)}  \, , \quad
\overline \pi^{(0)}_{ab} = \gamma^{(0)}_{ac} \, \gamma^{(0)}_{bd} \;
\overline \pi_{(0)}^{cd} \, .
\end{equation}
In eq.(\ref{dgcec}), the exponential of a matrix
$[A]$ is defined in terms of a Taylor series expansion,
\begin{equation}
{\rm exp} [A] = [I] + [A] + {1 \over 2!} [A]\, [A] + \ldots
\end{equation}
At this point it is important to remember that $[\chi(x), \gamma_{ab}(x)]$
and $[\gamma^{(0)}_{ab}(x), \pi_{(0)}^{ab}(x) ]$
are spatially dependent. Eq.(\ref{dgcea}-d) represents a general solution to
the strongly-coupled system consisting of gravity and dust.

\subsection{Relation to Kasner Metric}

Using semiclassical methods, Francisco and Pilati (1985) have shown how
to derive the Kasner metric describing pure gravity (see also
Salopek and Stewart 1993).
Here, it will be shown that the case of dust
interacting with gravity admits a solution where
the metric behaves like a Kasner metric at two separate epochs.

$(\gamma/\gamma_{(0)})^{1/2}$ is a quadratic polynomial in $\chi$,
\numparts
\begin{equation}
\left( \gamma/\gamma_{(0)} \right )^{1/2} =
\left( 1- { \chi \over \chi_{(1)}} \right )
\left( 1- { \chi \over \chi_{(2)}} \right ) \, ,
\end{equation}
with roots $(\chi_{(1)}, \chi_{(2)})$,
\begin{equation}
\chi_{(1)} = \left \{ {1\over2} \gamma_{(0)}^{-1/2} \pi_{(0)} +
	\sqrt{3\over2}
\left[ \gamma_{(0)}^{-1} \overline \pi^{ab}_{(0)} \overline \pi_{ab}^{(0)}
\right ]^{1/2}
	\right \}^{-1} \, ,
\label{root1}
\end{equation}
\begin{equation}
\chi_{(2)} = \left \{ {1\over2} \gamma_{(0)}^{-1/2} \pi_{(0)} -
	\sqrt{3\over2}
\left[ \gamma_{(0)}^{-1} \overline \pi^{ab}_{(0)} \overline \pi_{ab}^{(0)}
\right ]^{1/2}
	\right \}^{-1} \, ,
\label{root2}
\end{equation}
\endnumparts
which are spatially dependent: $\chi_{(1)} \equiv \chi_{(1)}(x)$ and
$\chi_{(2)} \equiv \chi_{(2)}(x)$.
The average of the two roots,
\begin{equation}
{ \chi_{(1)} + \chi_{(2)} \over 2 }
= { 2 \, \gamma_{(0)}^{1/2} \pi_{(0)} \over
\left( \pi_{(0)}^2
- 6 \overline \pi^{ab}_{(0)} \overline \pi_{ab}^{(0)} \right) }
\end{equation}
gives the time when the extrema of $(\gamma/\gamma_{(0)})^{1/2}$ is reached.
$z$ becomes
\begin{equation}
z = \sqrt{2\over 3} \ln  \left ( { 1 - {\chi \over \chi_{(2)}}  \over
			           1 - {\chi \over \chi_{(1)}} } \right )
   \, .
\end{equation}
Note that the equation for $[h]$ may be put in the following form
\begin{equation}
[h] = [h_{(0)}^{1/2}] \; {\rm exp}
\left \{
{2z \; [\gamma_{(0)}^{1/2}] \; [\overline \pi_{(0)}] \; [\gamma_{(0)}^{1/2}]
\over
\left( \overline \pi_{(0)}^{ab} \;
\overline \pi^{(0)}_{ab} \right )^{1/2} } \, \right \} \; [h_{(0)}^{1/2}] \, ,
\end{equation}
where the argument of the exponential is explicitly a symmetric matrix.
Following Salopek (1992), we diagonalize the argument using an orthogonal
matrix
$[O]$,
\numparts
\begin{equation}
{ [\gamma_{(0)}^{1/2}] \; [\overline \pi_{(0)}] \; [\gamma_{(0)}^{1/2}] \over
\left( \overline \pi_{(0)}^{ab} \;
\overline \pi^{(0)}_{ab} \right )^{1/2} } = \; [O]^T \; [D] \; [O] \, ,
\end{equation}
where
\begin{equation}
[O]^T \, [O] \, = \, [O] \, [O]^T = [I] \quad {\rm (identity \; matrix)} \, ,
\end{equation}
\begin{equation}
[D] = {\rm Diag} [d_1, d_2, d_3 ] \; , \quad
d_1 + d_2 + d_3 = 0 \, , \quad d^2_1 + d^2_2 + d^2_3 = 1 \, .
\end{equation}
\endnumparts
$[h]$ then admits the following simple form:
\begin{eqnarray}
[h] =  && [h_{(0)}^{1/2}] \, [O]^T \nonumber \\
&&{\rm Diag} \left [ \left ( { 1 - {\chi \over \chi_{(2)}}  \over
         1 - {\chi \over \chi_{(1)}} } \right )^{\sqrt{8 \over 3} d_1}  ,
\left ( { 1 - {\chi \over \chi_{(2)}}  \over
         1 - {\chi \over \chi_{(1)}} } \right )^{\sqrt{8 \over 3} d_2}  ,
\left ( { 1 - {\chi \over \chi_{(2)}}  \over
         1 - {\chi \over \chi_{(1)}} } \right )^{\sqrt{8 \over 3} d_3}
  \right ]  \\
&& [O] \; [h_{(0)}^{1/2}] \nonumber \,,
\end{eqnarray}
and $[\gamma]$
\begin{eqnarray}
[\gamma] =&&  [\gamma^{1/2}_{(0)}] \, [O]^T \,  \nonumber \\
{\rm Diag} \Bigg [
&& \left ( 1 - { \chi \over \chi_{(1)} } \right )^{2 p_1^{(1)}}
\left ( 1 - { \chi \over \chi_{(2)} } \right )^{2 p_2^{(1)}}, \nonumber\\
&& \left ( 1 - { \chi \over \chi_{(1)} } \right )^{2 p_1^{(2)}}
\left ( 1 - { \chi \over \chi_{(2)} } \right )^{2 p_2^{(2)}}, \nonumber\\
&& \left ( 1 - { \chi \over \chi_{(1)} } \right )^{2 p_1^{(3)}}
\left ( 1 - { \chi \over \chi_{(2)} } \right )^{2 p_2^{(3)}}  \Bigg ] \;
\label{kasner}\\
&& [O]\, [\gamma^{1/2}_{(0)}] \, , \nonumber
\end{eqnarray}
with
\numparts
\begin{equation}
p_1^{(i)} = { 1 \over 3 } - \sqrt{2 \over 3} d_i \, , \quad
p_2^{(i)} = { 1 \over 3 } + \sqrt{2 \over 3} d_i \, .
\end{equation}
Given a spatial point $x$, for times $\chi$ close to  $\chi_{(2)}(x)$,
the diagonal part of the 3-metric in eq.(\ref{kasner}) evolves like
a Kasner universe, with Kasner exponents,
$p_2^{(1)}, p_2^{(2)}, p_2^{(3)}$ :
\begin{equation}
p_2^{(1)} +  p_2^{(2)} + p_2^{(3)} = 1 \, , \quad
\left( p_2^{(1)} \right )^2  +  \left( p_2^{(2)} \right )^2  +
\left( p_2^{(3)} \right )^2 = 1  \, .
\end{equation}
\endnumparts
For times close to $\chi_{(1)}(x)$, the 3-metric also evolves like
a Kasner universe, but with Kasner exponents,
$p_1^{(1)}, p_1^{(2)}, p_1^{(3)}$. (See Belinski \etal 1970 and
Salopek and Stewart 1993.)

\subsection{Constant Functionals}

Given a generating functional ${\cal S}$ describing a HJ flow,
${\cal C}[\gamma_{ab}(x), \chi(x) ]$ is a
{\it Constant Functional} if for all choices of the lapse $N$ and
shift $N^i$, $\dot {\cal C}$ vanishes:
\begin{eqnarray}
0 = && \dot {\cal C} \equiv \int d^3x \;
\left [ { \delta {\cal C} \over \delta \chi(x) } \, \dot \chi(x) +
{ \delta {\cal C} \over \delta \gamma_{ab}(x) } \, \dot \gamma_{ab}(x)
\right ]  \\
= && \int d^3x \; \Bigg \{ {\delta {\cal C} \over \delta \chi(x)}
\left( N  + N^i \chi_{,i} \right ) + \nonumber \\
&& \left [
2 \, N \, \gamma^{-1/2}
\, \left [ 2 \gamma_{ac}(x) \gamma_{bd}(x) - \gamma_{ab}(x) \gamma_{cd}(x)
\right ] \; { \delta {\cal S} \over \delta \gamma_{cd}(x) }
+ N_{a|b} + N_{b|a}
\right ]
\,   { \delta {\cal C} \over \delta \gamma_{ab}(x) } \Bigg \} \, ,
\nonumber
\end{eqnarray}
where the last equality follows from the HJ flow equations,
\numparts
\begin{eqnarray}
\left( \dot \chi -N^i \chi_{,i} \right )/ N = &&1 \, , \\
\left( \dot \gamma_{ab} - N_{a|b} - N_{b|a} \right )/ N = &&
2 \gamma^{-1/2} \,
\left( 2 \gamma_{ac} \, \gamma_{bd} - \gamma_{ab} \, \gamma_{cd} \right ) \;
{ \delta {\cal S} \over \delta \gamma_{cd}(x) }   \, .
\end{eqnarray}
\endnumparts
Since $N$, $N_i$ are arbitrary, the constant functional ${\cal C}$ obeys the
following:
\numparts
\begin{equation}
{\delta {\cal C} \over \delta \chi(x)} +
2  \, \left [ 2 \gamma_{ac}(x) \gamma_{bd}(x) - \gamma_{ab}(x) \gamma_{cd}(x)
\right ] \; { \delta {\cal S} \over \delta \gamma_{ab}(x) } \;
{ \delta {\cal C} \over \delta \gamma_{cd}(x) } = 0 \, ,
\end{equation}
\begin{equation}
-2\left(\gamma_{ik} {\delta {\cal C} \over \delta \gamma_{kj} } \right)_{,j} +
{\delta {\cal C} \over \delta \gamma_{kl} } \, \gamma_{kl,i} +
{\delta {\cal C} \over \delta \chi } \, \chi_{,i} = 0 \, .
\end{equation}
\endnumparts

One may show that if
\begin{equation}
{\cal C} \equiv {\cal C}[\gamma^{(0)}_{ab}(x) ]
\label{constant}
\end{equation}
is an arbitrary gauge-invariant functional of the initial metric
$\gamma^{(0)}_{ab}(x)$ defined by the minimization prescription
eq.(\ref{minimize}), then
${\cal C}$ is indeed a constant functional. Two simple examples are,
\numparts
\begin{eqnarray}
{\cal C}_1 =&& \int d^3x \; \gamma_{(0)}^{1/2} \, , \\
{\cal C}_2 =&& \int d^3x \; \gamma_{(0)}^{1/2} \, R_{(0)} \, , \quad etc.
\end{eqnarray}
\endnumparts
Eq.(\ref{constant}) is an elegant result which has wide ranging implications.
For example, a complete list of constant functionals could
be used to define a particular universe within the
ensemble of universes whose evolution is described by the generating
functional ${\cal S}$. In fact, a space-time transformation of that particular
universe would not change the numerical values of the complete list
of constant functionals.

\section{Semiclassical Evolution: Dust and Gravity}

Solving the HJ equations, the energy constraint and the momentum constraint,
brings us one step closer to understanding the quantum theory of the
gravitational field because the generating functional ${\cal S}$,
eq.(\ref{cals}), may be interpreted as
the phase of the wavefunctional in the semiclassical approximation,
\begin{equation}
\Psi[ \gamma_{ab}(x), \chi(x)] \sim  e^{i {\cal S}[ \gamma_{ab}(x), \chi(x)]/
 \hbar} \, ,
\label{wavefunction}
\end{equation}
where Planck's constant $\hbar$ is assumed to be tiny. Explicit solutions for
${\cal S} \equiv {\cal S}[ \gamma_{ab}(x), \chi(x)]$,  are now discussed.

Semiclassical evolution is found by inverting the
classical evolution eqs.(\ref{dgcea}-d)
to express the initial metric,
$\gamma^{(0)}_{ab} \equiv \gamma^{(0)}_{ab}(\gamma_{cd}, \chi)$,
in terms of the original fields, and then
substituting to obtain ${\cal S}$ in eq.(\ref{cals}). Typically,
the inversion is very difficult because one must solve
nonlinear, partial differential equations. Some special
cases are given below and in Section 4 where the inversions may be
performed explicitly without resorting to an approximation method.
In the section  denoted {\it General Case}
below, one obtains semiclassical solutions by inverting the classical
evolution equations through  a Taylor series in $\chi(x)$.

\subsection{Elementary Example}
In the special case where the initial functional ${\cal Q}$
is proportional to the volume of the initial 3-geometry,
\begin{equation}
{\cal Q} = C \int d^3x \; \gamma_{(0)}^{1/2} \; ,
\quad {\rm (where \; C\; is \; a \; homogeneous\; constant) }
\end{equation}
it is easy to perform the necessary inversion to compute ${\cal S}$.
Since $\overline \pi^{ab}_{(0)}$ vanishes, one finds that $z=0$ and that
\numparts
\begin{eqnarray}
h^{(0)}_{ab} = && h_{ab}  \, , \\
\gamma_{(0)}^{1/2}
=&& {\gamma^{1/2} \over \left ( 1 - {3 \over 4} C \chi \right )^2 } \, .
\end{eqnarray}
\endnumparts
The generating functional ${\cal S}$ is then given by eq.(\ref{cals})
\begin{equation}
{\cal S}[\gamma_{ab}(x), \chi(x)] = -{4 \over 3} \int d^3x \, \gamma^{1/2} \,
{1 \over \left ( \chi(x) - {4 \over 3 C} \right ) } \, .
\end{equation}
This solution had be given previously by Salopek and Stewart (1992),
and it is of the form given in eq.(\ref{simple}).
It is mentioned here to show that the new formalism of this paper
encompasses the earlier work.

If $C$ is real, one may define the {\it classical Hubble parameter}, $H$,
through
\begin{equation}
H(x) \equiv - {\gamma^{-1/2}  \over 3} \; \gamma_{ab} \,
{ \delta {\cal S} \over \delta \gamma_{ab} } \, ,
\label{defHubble}
\end{equation}
giving
\begin{equation}
H(x) = {2 \over 3} \; {1 \over \left ( \chi(x) - {4 \over 3 C} \right ) } \, .
\label{defHubblex}
\end{equation}
Hence, if $C < 0$ and $\chi(x) \ge 0$, the Universe is expanding
(locally) at each spatial point,
whereas for $C>0$ and $\chi(x) \ge 0$, it is contracting locally until
it reaches a singularity at $\chi(x) = 4/(3C)$. The appearance of the
singularity is problematic for the classical theory.
$C=0$ yields the trivial solution ${\cal S}=0$.

\subsubsection{Evolution Beyond a Singularity}

Complex values of the parameter $C$ are of interest because they
help to explain how a universe may evolve beyond a singularity.
Let us write
\begin{equation}
{4 \over 3 C} = a - ib \, , {\rm where} \quad a, b \quad {\rm are\; real} \, ,
\end{equation}
and then split ${\cal S}$ into real and imaginary parts:
\numparts
\begin{equation}
{\cal S} = {\cal S}_R + i \, {\cal S}_I \, ,
\end{equation}
\begin{equation}
{\cal S}_R = - {4 \over 3 } \int d^3x \, \gamma^{1/2} \,
{ \left[ \chi(x) -a \right ] \over
\left[ \left( \chi(x) -a \right )^2 + b^2 \right ]   } \, ,
\end{equation}
\begin{equation}
{\cal S}_I =  {4 b \over 3 } \int d^3x \, \gamma^{1/2} \,
{ 1  \over
\left[ \left( \chi(x) -a \right )^2 + b^2 \right ]   } \, .
\end{equation}
\endnumparts
The wavefunctional then becomes
\begin{equation}
\Psi = {\rm exp} \left ( - {\cal S}_I / \hbar \right ) \;
       {\rm exp} \left ( i {\cal S}_R / \hbar \right )  \, .
\end{equation}
As a result, one can define a classical Hubble parameter using
the real part of the generating functional,
${\cal S}_R$, in eq.(\ref{defHubble}):
\begin{equation}
H(x) = {2 \over 3 } \;
{ \left[ \chi(x) -a \right ] \over
\left[ \left( \chi(x) -a \right )^2 + b^2 \right ]   }  \, .
\label{realhubble}
\end{equation}
(Here I am assuming that $H(x)$ or some closely related function
can be represented by some Hermitean
operator. In ordinary quantum mechanics, the expectation value
$<\hat p>$ of some Hermitean operator $\hat p$ is guaranteed to be
real. Hence for a wavefunction which is tightly peaked about a
classical trajectory, the expectation value of $\hat p$ may
be computed {\it approximately} using the {\it real} part of the phase.)

For $[ \chi(x) -a ]$ large and negative in eq.(\ref{realhubble}),
a universe is contracting (locally),
whereas for $[ \chi(x) -a ] $ large and positive,  the same universe is
expanding (locally). Apparently, that universe was able to pass through the
singularity at $\chi(x) = a$, and bounce from a contracting phase
to an expanding phase. If $b > 0$, universes with large volumes
are exponentially suppressed:
\begin{equation}
|\Psi|^2 = {\rm exp} \left \{ -{8b \over 3 \hbar} \;
\int d^3x \, \gamma^{1/2} \,
{ 1  \over
\left[ \left( \chi(x) -a \right )^2 + b^2 \right ]   }
\right \} \, .
\end{equation}
The above analysis was valid for arbitrary
$\chi(x)$. These qualitative results concerning the bounce from
a singularity are in agreement with a quantum formulation given by
Salopek (1992b) where he assumed that $\chi(x)$ was homogeneous.

\subsection{Intermediate Example}

The case of where the initial functional ${\cal Q}$
is a function of the initial volume is partially tractable:
\begin{equation}
{\cal Q} = f \left [ V_{(0)} \right ]  \,, \quad {\rm with} \quad
V_{(0)}= \int d^3x \, \gamma_{(0)}^{1/2} \, .
\end{equation}

After minimizing with respect to $\gamma^{(0)}_{ab}(x)$,
the classical evolution equations give
\numparts
\begin{equation}
z=0 \, ,   \quad h_{ab}(x) = h_{ab}^{(0)}(x) \, , \label{ssccea}
\end{equation}
\begin{equation}
\gamma^{1/2} = \left ( 1 - {3 f^\prime[V_{(0)}] \over 4 } \chi \right )^2 \;
\gamma_{(0)}^{1/2} \, , \label{sscceb}
\end{equation}
\endnumparts
where $f^\prime[V_{(0)}]$ denotes the derivative of the function
$f[V_{(0)}]$ with
respect to the variable $V_{(0)}$.
Hence, only the conformal factor of the 3-metric evolves.
Bringing the conformal factor to the other side, and integrating over
all $x$, one finds that, given a 3-metric $\gamma_{ab}(x)$ and
field configuration $\chi(x)$,  $V_{(0)}$ is defined
implicitly through
\numparts
\begin{equation}
V_{(0)} = \int d^3x \, \gamma^{1/2} \;
{ 1 \over \left[ 1 - {3 f^\prime[V_{(0)}] \over 4} \chi(x) \right ]^2 } \, .
\label{V0implicita}
\end{equation}
The resulting generating functional for arbitrary $\chi(x)$ is,
\begin{equation}
{\cal S} = -{4 \over 3} \int d^3x \, \gamma^{1/2} \,
{1 \over \left( \chi - {4 \over 3 f^\prime[V_{(0)}]} \right ) }
+ f[V_{(0)}] -V_{(0)} \,  f^\prime[V_{(0)}] \, .
\label{S2casea}
\end{equation}
\endnumparts
By invoking a Legendre transformation,
\numparts
\begin{equation}
g= f - {df \over d V_{(0)}} \, V_{(0)} \, ,
\end{equation}
\begin{equation}
b= {df \over d V_{(0)}} \, ,
\end{equation}
\endnumparts
one can simplify eqs.(\ref{V0implicita}-b) even further,
\numparts
\begin{equation}
{\cal S} = -{4 \over 3} \int d^3x \, \gamma^{1/2} \,
{1 \over \left( \chi - {4 \over 3 b}  \right ) } + g(b) \, ,
\label{S2caseb}
\end{equation}
\begin{equation}
{dg \over db}  = -
\int d^3x \, \gamma^{1/2} \;
{ 1 \over \left[ 1 - {3 b \over 4} \chi(x) \right ]^2 } \, .
\label{V0implicitb}
\end{equation}
\endnumparts
One verifies that the second eq.(\ref{V0implicitb})
is a consequence of minimizing the first eq.(\ref{S2caseb})
with respect to the single independent variable $b$.
Finally, using one more set of substitutions,
\begin{equation}
a= {4 \over 3 b} \, , \quad j(a) = g(b)
\end{equation}
one can express the generating functional in its simplest form,
\numparts
\begin{equation}
{\cal S}[\gamma_{ab}(x), \chi(x)] = -{4 \over 3} \int d^3x \, \gamma^{1/2} \,
{1 \over \left[ \chi(x) - a  \right ] } + j(a) \, ,
\label{S2casec}
\end{equation}
\begin{equation}
0 = - {4 \over 3 } \, \int d^3x \, \gamma^{1/2} \;
{ 1 \over \left[ \chi(x)  - a \right ]^2 } + {dj \over da }\, .
\label{aimplicit}
\end{equation}
\endnumparts
Once again, the second eq.(\ref{aimplicit}) is obtained from the
first by minimizing ${\cal S}$  with respect to $a$ where
$j(a)$ is an arbitrary function of $a$.

One could have derived this result more simply
by starting with the solution,
\begin{equation}
{\cal T}[\gamma_{ab}, \chi(x)| a] =
-{4 \over 3} \int d^3x \, \gamma^{1/2} \,
{1 \over \left[ \chi(x) - a  \right ] }
\end{equation}
of the energy constraint {\it and} the momentum constraint,
and by then constructing another solution ${\cal S}$ by taking
a superposition over the homogeneous parameter $a$,
\numparts
\begin{equation}
{\cal S}[\gamma_{ab}(x), \chi(x)] = {\cal T}[\gamma_{ab}, \chi(x)| a]
+ j(a) \, ,
\end{equation}
which means that one chooses $a$ according to the minimization
prescription:
\begin{equation}
0 = {\partial {\cal T} \over \partial a} + { dj \over da} \, .
\end{equation}
\endnumparts
This solution was actually suggested by Salopek and Bond (1990)
who gave a much different derivation.
Here the aim was to show that the Green's function method encompasses
previous results. In addition, the Green's function method has the
advantage that it yields the classical evolution equations
eq.(\ref{ssccea}-b).

\subsubsection{Explicit Computation for Intermediate Example}

If the arbitrary function $j(a)$ is assumed to be linear,
\numparts
\begin{equation}
j(a) = a \, ,
\end{equation}
one may readily compute the generating functional eq.(\ref{S2casec}),
if $\chi(x) = \overline \chi$ is homogeneous,
\begin{equation}
{\cal S}[\gamma_{ab}(x), \chi(x) = \overline \chi] =
\overline \chi + \left [ {16 \over 3 } \int d^3x \, \gamma^{1/2} \right ]^{1/2}
\quad {\rm (exact)} \, ,
\label{Schih}
\end{equation}
since one may invert eq.(\ref{aimplicit}) to give,
\begin{equation}
a = \overline \chi + \left [ {4 \over 3 } \int d^3x \,
\gamma^{1/2} \right ]^{1/2}
\quad {\rm (exact)} \, .
\label{asolution}
\end{equation}
\endnumparts
The sign before the square root in eq.(\ref{asolution}) is arbitrary,
and I have chosen it to be positive. (Choosing a negative sign yields another
solution of the HJ equation.)

One may obtain an approximate form which is valid
for mildly inhomogeneous fields $\chi(x)$, by assuming that
$a$ is large, and then expanding the implicit equation (\ref{aimplicit})
in powers of $\Delta \chi(x)/a$ where
\begin{equation}
\Delta \chi(x)  = \chi(x) - \overline \chi \, , \quad
\overline \chi = {1\over V} \, \int d^3 x \, \gamma^{1/2} \, \chi(x) \, ,
\quad   V = \int d^3 x \, \gamma^{1/2} \, .
\end{equation}
The first few terms of the implicit equation (\ref{aimplicit}) yield,
\begin{equation}
a = \overline \chi + \sqrt{4V\over 3}
\left[ 1 + {9 \over 8 V} <(\chi -\overline \chi)^2>  \right ]
+ \ldots \, .
\end{equation}
The generating functional then becomes
\numparts
\begin{equation}
{\cal S} = \sqrt{ 16 V \over 3 } + \overline \chi
+ \sqrt{ 3 \over 4 V} \, < \left ( \chi - \overline \chi \right )^2 >
+ \ldots \, ,
\end{equation}
with
\begin{equation}
<\left( \chi - \overline \chi \right )^2> =
{1\over V} \, \int d^3x \, \gamma^{1/2} \,
\left( \chi(x)- \overline \chi \right )^2 \, .
\end{equation}
\endnumparts
If $\chi(x)= \overline \chi$ is homogeneous, then one recovers the earlier
result, eq.(\ref{Schih}).

\subsection{General Case}

It is straightforward to write down the generating functional for
${\cal S}$ in terms of $\chi(x)$ and the initial 3-metric:
\numparts
\begin{equation}
{\cal S} = \int d^3x \, \gamma_{(0)}^{-1/2} \, \chi(x) \,
\left \{ 2 \overline \pi_{(0)}^{ab} \, \overline \pi^{(0)}_{ab} -
{1\over 3}  \, \pi_{(0)}^2 \right \} + Q[\gamma^{(0)}_{ab}] \label{tfunc} \, ,
\end{equation}
\begin{equation}
\pi^{ab}_{(0)} = { \delta {\cal Q} \over \delta \gamma_{ab}^{(0)}(x) } \, .
\end{equation}
\endnumparts
Using the roots
$( \chi_{(1)}, \chi_{(2)} )$ of $(\gamma/\gamma_{(0)})^{1/2}$ defined in
eqs.(\ref{root1}-c), one
may rewrite the generating functional in the elegant form:
\begin{equation}
{\cal S} = - {4 \over 3} \, \int d^3x \, \gamma_{(0)}^{1/2} \;
{\chi \over \chi_{(1)} \, \chi_{(2)} } \, + {\cal Q}[\gamma^{(0)}_{ab}].
\label{elegant}
\end{equation}

For some applications, one would like to express this as a function
of the original 3-metric $\gamma_{ab}(x)$.  In general, the necessary
inversion of eqs.(\ref{dgcea}-\ref{dgced}) is very difficult. However,
in the limit as $\chi(x) \rightarrow 0$, this inversion is straightforward
since
\begin{equation}
\gamma_{ab}(x) \rightarrow \gamma^{(0)}_{ab}(x)
\quad {\rm as} \; \; \chi(x) \rightarrow 0 \, .
\end{equation}
Hence from eq.(\ref{tfunc}), one concludes that
\begin{equation}
{\cal S} \rightarrow {\cal Q}
\quad {\rm as} \; \; \chi(x) \rightarrow 0 \, .
\end{equation}
which justifies the claim made earlier in eq.(\ref{initialstate}).

\subsection{Advanced Example}

In order to be concrete, I will illustrate the main features of
the {\it General Case} by assuming that
\numparts
\begin{equation}
{\cal Q} = \int d^3 x \; \gamma_{(0)}^{1/2} \; \left [ C + E R_{(0)} \right ]
\, ,
\end{equation}
in which case
\begin{equation}
\gamma_{(0)}^{-1/2} \pi_{(0)} = {1\over 2} \left( 3C + E R_{(0)} \right ) \,
\quad {\rm and} \quad \gamma_{(0)}^{-1/2} \overline \pi_{(0)}^{ab} =
- E \overline R^{ab} \, .
\end{equation}
\endnumparts

\subsubsection{Early Time Behavior}

Expanding the classical solutions (\ref{dgcea}-c) in a Taylor series in
$\chi(x)$,
one finds
\numparts
\begin{equation}
\left(\gamma / \gamma_{(0)} \right)^{1/2} = 1 - {1\over 2}
\left( 3C + R_{(0)} \right ) \, \chi  + \ldots \, ,
\end{equation}
\begin{equation}
z = 2 E \, \sqrt{ \overline R_{(0)}^{ab} \overline R^{(0)}_{ab} } \,
\chi + \ldots \, ,
\end{equation}
\begin{equation}
[h] = [h^{(0)}] \left ( 1 - 4 E [\overline R_{(0}] [ \gamma^{(0)} ] \, \chi
+ \ldots \right ) \, ,
\end{equation}
\endnumparts
where $[ \overline R_{ (0) } ]$ denotes the traceless Ricci tensor with
contravariant indices:
\begin{equation}
[\overline R_{(0)}]^{ab} \equiv R_{(0)}^{ab}
- {1\over 3 } \gamma_{(0)}^{ab}\, R_{(0)}
   \, .
\end{equation}
Only first order is shown here although one could expand to arbitrary order.
Using an iterative process, one can invert these equations to obtain
$\gamma^{(0)}_{ab}$ as a function of $\chi$ and $\gamma$,
\begin{equation}
\gamma^{(0)}_{ab}(x) = \gamma_{ab}(x) + \chi(x) \, \left \{
C \gamma_{ab}(x)  + E \left [ 4 R_{ab}(x) - R(x) \, \gamma_{ab}(x) \right ]
\right \}
+ \ldots
\end{equation}
Substituting into eq.(\ref{tfunc}), ${\cal S}$ becomes
\begin{eqnarray}
{\cal S} =&& \int d^3x \, \gamma^{1/2} \, \left ( C + E R \right )
+ \nonumber \\
&& \int d^3 x \, \gamma^{1/2} \, \chi(x) \, \left [
E^2 \left( {3 \over 4} R^2 - 2 R^{ab} R_{ab} \right ) +
{1\over 2} C E R + {3\over 4} C^2 \right ]
+\ldots \, .
\end{eqnarray}
This agrees with a similar calculation by Salopek (1997).

\subsubsection{Late Time Behavior}

In order to determine the late time behavior of ${\cal S}$ ,
one introduces a factor of $\gamma^{1/2}$ in eq.(\ref{tfunc}):
\begin{equation}
{\cal S} = \int d^3x \, \gamma^{1/2} \, \chi \,
{ \left [ 2 \overline \pi_{(0)}^{ab} \, \overline \pi^{(0)}_{ab} -
{1\over 3}  \, \pi_{(0)}^2 \right ]  \over
\left [ \left( 1 - {1 \over 2}  \, \chi \,
\gamma_{(0)}^{-1/2} \pi_{(0)} \right )^2
- {3\over2 } \chi^2 \gamma_{(0)}^{-1}
\, \overline \pi_{(0)}^{ab} \, \overline \pi^{(0)}_{ab}
\right ] }
+ Q[\gamma^{(0)}_{ab}] \label{tfunclate} \, ,
\end{equation}
In the limit that $\chi$ is large and positive, ${\cal S}$ assumes
the very simple form,
\begin{equation}
{\cal S}[\gamma_{ab}(x), \chi(x)] =
- {4 \over3 } \int d^3 x \, \gamma^{1/2} \, { 1\over \chi(x)} \, .
\end{equation}
However, it may not always be possible to take the limit to large $\chi$
because one may first reach a singularity at a finite value of $\chi$.
For this case, the form of the generating functional in terms of the
original variables has not as yet been determined.

\section{Strongly-Coupled Solutions: Two Dust Fields Interacting with Gravity}

A non-trivial example which illustrates
the main features of the Green's function solution method is provided
by two dust fields, $\chi_1$ and $\chi_2$,
interacting with gravity. The constraint equations for the
long-wavelength system are:
\numparts
\begin{equation}
{\cal H}^{(s)}(x)/ \kappa = \gamma^{-1/2} \,
\left ( 2 \gamma_{ac} \gamma_{bd} - \gamma_{ab} \gamma_{cd} \right )
{\delta {\cal S} \over \delta \gamma_{ab}}
{\delta {\cal S} \over \delta \gamma_{cd}} +
{\delta {\cal S} \over \delta \chi_1} +
{\delta {\cal S} \over \delta \chi_2}  = 0 \, ,
\label{hjddg}
\end{equation}
\begin{equation}
{\cal H}^{(s)}_{i}(x)=
-2\left(\gamma_{ik}
{\delta {\cal S} \over \delta \gamma_{kj} } \right)_{,j} +
{\delta {\cal S} \over \delta \gamma_{kl} } \, \gamma_{kl,i} +
{\delta {\cal S} \over \delta \chi_1 } \, \chi_{1,i} +
{\delta {\cal S} \over \delta \chi_2 } \, \chi_{2,i} = 0 \, .
\label{mcddg}
\end{equation}
\endnumparts
After defining the {\it average field}, $\chi(x)$, and the {\it static field},
$y(x)$, through
\begin{equation}
\chi = { (\chi_2 + \chi_1 ) \over 2} \, , \quad
y = { (\chi_2 - \chi_1 ) \over 2} \; ,
\end{equation}
the constraints become
\numparts
\begin{equation}
{\cal H}^{(s)}(x)/ \kappa =\gamma^{-1/2} \,
\left ( 2 \gamma_{ac} \gamma_{bd} - \gamma_{ab} \gamma_{cd} \right )
{\delta {\cal S} \over \delta \gamma_{ab}}
{\delta {\cal S} \over \delta \gamma_{cd}} +
{\delta {\cal S} \over \delta \chi}   = 0 \, ,
\label{hjddgp}
\end{equation}
\begin{equation}
{\cal H}^{(s)}_{i}(x)=
-2\left(\gamma_{ik}
{\delta {\cal S} \over \delta \gamma_{kj} } \right)_{,j} +
{\delta {\cal S} \over \delta \gamma_{kl} } \, \gamma_{kl,i} +
{\delta {\cal S} \over \delta \chi } \, \chi_{,i} +
{\delta {\cal S} \over \delta y } \, y_{,i} = 0 \, . \label{mcddgp}
\end{equation}
\endnumparts
The static field $y(x)$ appears in the momentum constraint eq.(\ref{mcddgp})
but it is absent in the energy constraint eq.(\ref{hjddgp}) which is
identical to that of a single dust field, eq.(\ref{hjsdg}). For all intents,
one can describe the two-dust system with gravity, which consists of the
fields $(\chi_1, \chi_2, \gamma_{ab})$, as a system containing
a single dust field
interacting with gravity, $\chi$ and $\gamma_{ab}$,
and containing a static field
$y(x)$ which appears in the generating functional ${\cal S}$ but which
does not evolve. Hence given the initial functional,
\begin{equation}
{\cal Q} \equiv {\cal Q}[\gamma_{ab}^{(0)}(x), y(x)] \,
\end{equation}
which is invariant upon reparametrization of the spatial coordinates,
\begin{equation}
-2\left(
\gamma^{(0)}_{ik} \, { \delta {\cal Q} \over \delta \gamma^{(0)}_{kj} }
\right)_{,j} +
{ \delta {\cal Q} \over \delta \gamma^{(0)}_{kl} } \,
\gamma^{(0)}_{kl,i} +
{ \delta {\cal Q} \over \delta y }  \, y_{,i} = 0 \, , \label{mcddgpp}
\end{equation}
the generating functional at later times is given by
\numparts
\begin{equation}
{\cal S}[\gamma_{ab}, \chi(x), y(x)] =
{\cal G}[\gamma_{ab}(x), \chi(x)| \gamma^{(0)}_{ab}(x), \chi_{(0)}(x) = 0]
+ {\cal Q}[\gamma^{(0)}_{ab}(x), y(x)]  \,
\end{equation}
where ${\cal G}$ is just the Green's function for dust with gravity defined in
eq.(\ref{greenf}) with $\chi_{(0)}(x) = 0$, and
$\gamma^{(0)}_{ab}(x)$ chosen to minimize ${\cal G} + {\cal Q}$:
\begin{equation}
0 = {\delta {\cal G} \over \delta \gamma^{(0)}_{ab} }  +
{\delta {\cal Q} \over \delta \gamma^{(0)}_{ab} } \, .
\end{equation}
\endnumparts
The minimization prescription leads to the classical
evolution equations (\ref{dgcea}-d) which carry over
directly to the present situation.

\subsection{Exact Solution for Two Dust Fields with Gravity}

Explicit solutions arise when the initial generating functional is  of the
form
\begin{equation}
{\cal Q}[\gamma^{(0)}_{ab}(x), y(x)] = \int d^3x \, \gamma_{(0)}^{1/2} \,
\left [ A + \left( \gamma_{(0)}^{ab} \, y_{,a} y_{,b} \right )^{1/2}
\right ] \, , \label{iddg}
\end{equation}
where $A$ is a constant.
Using the Green's function method described above,
I will show in section 4.2 that
the resulting generating functional ${\cal S}$ at arbitrary
$\chi(x)$ is
\begin{eqnarray}
{\cal S}[\gamma_{ab}(x), \chi(x), y(x)] =&&
-{4 \over 3} \int d^3x \, \gamma^{1/2} \,
{1 \over \left ( \chi - {4 \over 3 A} \right )  }
+ \nonumber \\
&&\int d^3x \, \gamma^{1/2} \,
\left( 1 - {3A \over 4} \chi \right )^{-4/3} \,
\left ( \gamma^{ab} y_{,a} y_{,b} \right )^{1/2} \, . \label{sddg}
\end{eqnarray}
This solution is a special case considered by Salopek, Stewart and
Parry (1993).
One may also demonstrate the validity of this solution by direct
substitution into the HJ eq.(\ref{hjddgp}).

\subsection{Derivation of Exact Solution}

The explicit solution (\ref{sddg}) for two dust fields interacting with
gravity will be justified using the Green's function method of Section 2.

The momentum conjugate to $\gamma^{(0)}_{ab}$ is given by eq.(\ref{dgced}):
\numparts
\begin{equation}
\gamma_{(0)}^{-1/2}  \pi_{(0)}^{ab} =
{1 \over 2  } \; \left [
\gamma_{(0)}^{ab} \; ( A + s^{1/2} ) - s^{-1/2} \, y^{;a} \, y^{;b}
\right ] \, ,
\end{equation}
where
\begin{equation}
s = \gamma_{(0)}^{ab} \; y_{,a} y_{,b} \,
\quad {\rm and} \quad y^{;a} \equiv \gamma^{ab}_{(0)} \; y_{,b} \, .
\label{sdef}
\end{equation}
\endnumparts
This example is qualitatively different from the exact solution given
in Section 3 in that the traceless part of $\pi_{(0)}^{ab}$
is non-vanishing,
\numparts
\begin{equation}
\gamma_{(0)}^{-1/2} \, \overline \pi_{(0)}^{ab} =
- {1 \over 2 s^{1/2} } \left [ y^{;a} y^{;b} - {s \over 3} \gamma_{(0)}^{ab}
\right ]  \, ,
\end{equation}
while the trace is given by
\begin{equation}
\gamma_{(0)}^{-1/2} \, \pi_{(0)} = {3 A \over 2} + s^{1/2} \, .
\end{equation}
\endnumparts
The square root of the determinant $\gamma$ evolves as a quadratic
function of $\chi$,
\numparts
\begin{equation}
\left (\gamma/ \gamma_{(0)} \right )^{1/2} =
\left ( 1 - { \chi \over \chi_{(1)} } \right ) \,
\left ( 1 - { \chi \over \chi_{(2)} } \right ) \, ,
\end{equation}
with roots at
\begin{equation}
\chi_{(1)} = {1 \over {3A \over 4} + s^{1/2} } \, ,
\end{equation}
\begin{equation}
\chi_{(2)} = { 4 \over 3 A} \, .
\end{equation}
\endnumparts
Please note that $\chi_{(1)} \ne \chi_1$; $\chi_{(1)}$ denotes
the first root of $( \gamma/ \gamma_{(0)} )^{1/2}$ whereas
$\chi_1$ denotes the first dust field.  The 3-metric evolves according to
\numparts
\begin{equation}
\gamma_{ab} = \left( 1 - { \chi \over \chi_{(2)} } \right )^{4/3} \,
B_{ac}  \; \gamma_{(0)}^{cd} \; B_{db}  \, ,
\end{equation}
where the symmetric matrix $B_{ac}$ has components
\begin{equation}
B_{ac} = \gamma^{(0)}_{ac} - {1\over s } \,
{ \left ( { \chi \over \chi_{(1)} } - { \chi \over \chi_{(2)} }  \right )
\over \left ( 1 - { \chi \over \chi_{(2)} }  \right ) } \,
y_{,a} \, y_{,c} \, .
\end{equation}
\endnumparts
The inverse 3-metric has components
\numparts
\begin{equation}
\gamma^{ab} =
\left( 1 - { \chi \over \chi_{(2)} } \right )^{-4/3}  \,
C^{ac} \, \gamma^{(0)}_{cd} \, C^{db} \, .
\end{equation}
where $C^{ac}$ is a symmetric matrix:
\begin{equation}
C^{ac} = \gamma_{(0)}^{ac} +
{1\over s} \,
{ \left ( { \chi \over \chi_{(1)} } - { \chi \over \chi_{(2)} }  \right )
\over \left ( 1 - { \chi \over \chi_{(1)} } ) \right ) } \,
y^{;a} \, y^{;c} \, .
\end{equation}
\endnumparts
One concludes that
\begin{equation}
y^{|a} \, y_{|a} \equiv \gamma^{ab} \, y_{,a} \, y_{,b} =
{ \left ( 1 - { \chi \over \chi_{(2)} } \right )^{2/3}
\over  \left ( 1 - { \chi \over \chi_{(1)} } \right )^2 } \;
\gamma_{(0)}^{ab} \, y_{,a} \, y_{,b} \, .
\end{equation}
One may readily invert this relation to find
$s$, eq.(\ref{sdef}), as function of the original fields:
\begin{equation}
s^{1/2} = {
    \left ( 1 - {3A \over 4} \chi \right ) \over
    		\left [
\chi +  \left ( 1 - {3A \over 4} \chi \right )^{1/3}
    \left( y^{|a} \, y_{|a} \right )^{-1/2}
		\right ]
	  } 	 \, .
\end{equation}
Using eq.(\ref{elegant}), one can write ${\cal S}$
in terms of $\chi(x)$ and the initial 3-metric,
\begin{equation}
{\cal S} = \int d^3x \, \gamma_{(0)}^{1/2} \left [
A \left( 1 - {3A \over 4} \chi \right ) + s^{1/2} ( 1 - A \, \chi )
\right ]
\end{equation}
Eliminating the initial variables in favor of the original ones, one
recovers the claimed result eq.(\ref{sddg}).

\section{Strongly-Coupled Solutions: Scalar Field Interacting with Gravity}

\subsection{Massless Scalar Field, Gravity and Cosmological Constant}

The strongly-coupled system of gravity interacting with a scalar field is also
tractable. Most of the techniques that applied to dust and gravity can also be
applied here although the details are quite different. The main results will
be stated without much elaboration.

In the strongly-coupled limit, the Hamiltonian density
for gravity and a scalar field with cosmological constant is:
\begin{equation}
{\cal H}^{(s)}(x)/ \kappa =
\gamma^{-1/2} \, \left ( 2 \gamma_{ac} \gamma_{bd}
- \gamma_{ab} \gamma_{cd} \right )
{\delta {\cal S} \over \delta \gamma_{ab}}
{\delta {\cal S} \over \delta \gamma_{cd}} +
{1\over 2} \gamma^{-1/2} \,
\left( {\delta {\cal S} \over \delta \phi}  \right )^2 +
\gamma^{1/2} \, V_0= 0 \, .
\label{hjsg}
\end{equation}

A complete solution (``Green's function solution'') for the generating
functional describing a massless scalar
field with cosmological constant term interacting with gravity is:
\begin{eqnarray}
{\cal G}[\gamma_{ab}(x), &&\phi(x)| \; \gamma^{(0)}_{ab}(x), \phi_{(0)}(x) ] =
- \sqrt{ 4 V_0 \over 3 } \int d^3 x \; \nonumber \\
&&\left[ \gamma+ \gamma_{(0)} - 2 \gamma^{1/2} \gamma_{(0)}^{1/2}
{\rm cosh}
\left ( \sqrt{3 \over2} \sqrt{ z^2 + (\phi - \phi_{(0)} )^2 } \right )
\right ]^{1/2} \; ,
\label{greenmc}
\end{eqnarray}
where $z$ was defined in eq.(\ref{defz}).
The sign of the Green's function is arbitrary and a minus sign was chosen
in eq.(\ref{greenmc}).
Hereafter, I will assume that $\phi_{(0)}= 0$ for similar
reasons that were given for a dust field in Sections 2.1 and 2.2.

In order to satisfy the momentum constraint,
\begin{equation}
{\cal H}^{(s)}_{i}(x)=-2\left(\gamma_{ik} \,
{\delta {\cal S} \over \delta \gamma_{kj}} \right)_{,j} +
{\delta {\cal S} \over \delta \gamma_{kl}} \gamma_{kl,i}
+ { \delta {\cal S} \over \delta \phi } \, \phi_{,i} = 0 \, ,
\end{equation}
one constructs a solution ${\cal S}$ using the Superposition Principle:
\numparts
\begin{equation}
{\cal S}[\gamma_{ab}(x), \phi(x)] =
{\cal G}[\gamma_{ab}(x), \phi(x)| \; \gamma^{(0)}_{ab}(x),  \phi_{(0)}(x)=0] +
{\cal Q}[\gamma^{(0)}_{ab}] \, ,
\end{equation}
where $\gamma^{(0)}_{ab}(x)$ has been chosen to minimize
${\cal G} \; + \; {\cal Q}$,
\begin{equation}
0 = {\delta {\cal G} \over \delta \gamma^{(0)}_{ab} }  +
{\delta {\cal Q} \over \delta \gamma^{(0)}_{ab} } \, .
\end{equation}
\endnumparts
${\cal Q}$ is an arbitrary gauge-invariant functional of $\gamma^{(0)}_{ab}$.
Please note that ${\cal S}$ coincides with ${\cal Q}$ when $\phi(x) = 0$:
\begin{equation}
{\cal S}[ \gamma_{ab}(x), \phi(x)=0] = {\cal Q}[\gamma_{ab}(x)] \, .
\end{equation}

\subsection{Classical Evolution}

Classical evolution is given by:
\numparts
\begin{equation}
\left( \gamma / \gamma_0 \right )^{1/2} =
{ 1 \over {\rm cosh} \theta + { A \over \sqrt{ A^2 - 1 } } {\rm sinh} \theta }
\, , \label{sgcc}
\end{equation}
\begin{equation}
z =  \phi \;  \Bigg /
\sqrt{  { \left( {1 \over 3} \gamma^{-1}_{(0)} \pi_{(0)}^2 - V_0 \right )
\over 2 \gamma_{(0)}^{-1} \overline \pi_{(0)}^{ab} \;
\overline \pi^{(0)}_{ab} } -1 } \, , \label{sgcd}
\end{equation}
\begin{equation}
[h] = [h^{(0)}] \; {\rm exp}
\left [
{2z \; [\overline \pi_{(0)}] \; [\gamma^{(0)}] \over
\left( \overline \pi_{(0)}^{ab} \;
\overline \pi^{(0)}_{ab} \right )^{1/2} } \, \right ] \; .
\label{sgce}
\end{equation}
\endnumparts
$A$ and $\theta$ are simply abbreviations for the following expressions:
\numparts
\begin{equation}
A = { \gamma_{(0)}^{-1/2} \pi_{(0)} \over \sqrt{3 V_0} } \, ,
\label{sgca}
\end{equation}
\begin{equation}
\theta = \sqrt{3\over2} \, \phi \;
\sqrt{  {1 \over 3 } \gamma_{(0)}^{-1} \, \pi_{(0)}^2 - V_0 } \;
\Bigg /
\sqrt{ {1\over 3} \gamma_{(0)}^{-1} \, \pi_{(0)}^2
- 2 \gamma_{(0)}^{-1}
\overline \pi_{(0)}^{ab} \; \overline \pi^{(0)}_{ab}
-  V_0  }\, ,
\label{sgcb}
\end{equation}
\endnumparts

\section{Semiclassical Evolution: Scalar Field,  Gravity and
Cosmological Constant}

It is straightforward to write ${\cal S}$ in terms of $\phi(x)$ and
the initial 3-metric,
\begin{equation}
{\cal S} = - \sqrt{4 V_0 \over 3} \,
\int d^3x \, \gamma_{(0)}^{1/2} \;
{ {\rm sinh} \theta \over
\left( \sqrt{A^2 - 1} \, {\rm cosh}\theta + A {\rm sinh} \theta \right ) }
+ {\cal Q}[\gamma_{ab}^{(0)}] \, , \label{scalarnew}
\end{equation}
with $A$ and $\theta$ defined in eqs.(\ref{sgca},b),
but it is very difficult to find an explicit expression in terms of the
original variables. I will be content to illustrate the `early' and
`late' time behavior.

\subsection{Behavior for small $\phi(x)$ }

To this aim,  note that for small $\phi(x)$
the original 3-metric evolves according to
\begin{equation}
\gamma_{ab} = \gamma^{(0)}_{ab}  +
{ \sqrt{2} \; \phi \; \gamma_{(0)}^{-1/2} \over
\sqrt{ {1\over 3} \gamma_{(0)}^{-1} \, \pi_{(0)}^2
- 2 \gamma_{(0)}^{-1} \overline \pi_{(0)}^{ab} \; \overline \pi^{(0)}_{ab}
-  V_0  } } \;
\left[ 2 \pi_{ab}^{(0)} - \pi_{(0)} \, \gamma^{(0)}_{ab}  \right ] + \dots
\end{equation}
This may be inverted by iteration to give $\gamma^{(0)}_{ab}(x)$ as
a function of $\phi(x)$ and $\gamma_{ab}(x)$:
\begin{equation}
\gamma^{(0)}_{ab}  =  \gamma_{ab} -
{ \sqrt{2} \;  \phi \; \gamma^{-1/2}
\over
\sqrt{ {1\over 3} \gamma^{-1} \, \pi^2
- 2 \gamma^{-1} \overline \pi^{ab} \; \overline \pi_{ab}
-  V_0 } } \;
\left[ 2 \pi_{ab} - \pi \, \gamma_{ab}  \right ] + \dots \, .
\label{inversion}
\end{equation}
Substituting into eq.(\ref{scalarnew}), one finds to first order in
$\phi(x)$ that
\numparts
\begin{eqnarray}
{\cal S}[&& \gamma_{ab}(x), \phi(x) ]=  {\cal Q}[\gamma_{ab}(x) ] + \nonumber\\
&& \sqrt{2} \, \int d^3x \, \gamma^{1/2} \, \phi \;
\sqrt{ {1\over 3} \gamma^{-1} \, \pi^2
- 2 \gamma^{-1} \overline \pi^{ab} \; \overline \pi_{ab}
-  V_0  } + \ldots \, . \\
&&( {\rm small} \; \; \phi(x) \; \; {\rm behavior } )  \nonumber
\end{eqnarray}
with
\begin{equation}
\pi^{ab} = {\delta {\cal Q} \over \delta \gamma_{ab}  } \, , \quad
\pi = \gamma_{ab} \, \pi^{ab} \, , \quad {\rm and} \quad
\overline \pi^{ab} = \pi^{ab} - {1\over 3 } \pi \, \gamma^{ab} \, ,
\end{equation}
\endnumparts
which is in agreement with Salopek (1997).

\subsection{Behavior for large $\gamma^{1/2}(x)$ }

In order to determine the behavior at large $\gamma^{1/2}$,
one rewrites the generating functional ${\cal S}$ as
\begin{equation}
{\cal S} = - \sqrt{ 4 V_0 \over 3} \,
\int d^3x \, \gamma^{1/2} \left [
1+ {\gamma_{(0)} \over \gamma} - 2
\left( \gamma_{(0)} \over \gamma \right)^{1/2} \,
{\rm cosh} \theta \right ]^{1/2} + {\cal Q}[\gamma^{(0)}_{ab}] \, .
\end{equation}
Provided $A^2 -1 > 0$,
$\gamma$ becomes very large as $\theta \rightarrow \theta_{crit}$
where $\theta_{crit}$ is given by
\begin{equation}
\theta_{crit} = - {\rm tanh}^{-1}
\left ( { \sqrt{ A^2 -1 } \over A } \right ) \, .
\end{equation}
In this limit,
$\gamma_{(0)}/ \gamma \rightarrow 0$, and ${\cal S}$
is proportional to the volume of any given 3-geometry,
\begin{equation}
{\cal S}[\gamma_{ab}(x), \phi(x)] =
- \sqrt{ 4 V_0 \over 3} \, \int d^3x \, \gamma^{1/2} \, ,
\quad ({\rm large} \; \; \gamma^{1/2}(x) \; \; {\rm behavior}) \, .
\end{equation}
For the sake of brevity, other cases will not be considered here.

\section{Massless Scalar Field and Gravity}

A complete solution for the generating functional describing gravity and
a massless scalar field without cosmological constant is:
\begin{equation}
{\cal G}[\gamma_{ab}(x), \phi(x)| \; \gamma^{(0)}_{ab}(x), \phi_{(0)}(x) ] =
- \int d^3x  \; \gamma^{1/4}  \; \gamma_{(0)}^{1/4} \;
{\rm exp} - \sqrt{3 \over 8} \sqrt{ z^2 + (\phi - \phi_{(0)} ) ^2 } \; ,
\end{equation}
where $z$ is still defined by eq.({\ref{defz}).
I will assume again that $\phi_{(0)}=0$.

\subsection{Classical Evolution}

Classical evolution is given by
\numparts
\begin{equation}
\left( { \gamma \over \gamma_{(0)} } \right )^{1/4}  =
{4 \over 3} \; \gamma_{(0)}^{-1/2}\, \pi_{(0)} \;
{\rm exp} \left [ \sqrt{ 3 \over 8 }
{ \pi_{(0)} \, \phi \over \sqrt{ \pi_{(0)}^2
- 6 \overline \pi^{ab}_{(0)} \, \overline \pi_{ab}^{(0)} } } \right ] \; ,
\end{equation}
\begin{equation}
z = { \phi \over
\sqrt{ { \pi_{(0)}^2 \over 6 \overline \pi^{ab}_{(0)} \,
\overline \pi_{ab}^{(0)} }
- 1 } } \, ,
\end{equation}
\begin{equation}
[h] = [h^{(0)}] \; {\rm exp}
\left [
{2z \; [\overline \pi_{(0)}] \; [\gamma^{(0)}] \over
\left( \overline \pi_{(0)}^{ab} \;
\overline \pi^{(0)}_{ab} \right )^{1/2} } \, \right ] \; .
\end{equation}
\endnumparts
In particular for
\numparts
\begin{equation}
{\cal Q} = \int d^3x \; \gamma_{(0)}^{1/2} \; \left [ C + E R_{(0)} \right ]
\, ,
\end{equation}
the classical evolution is given by,
\begin{equation}
\left( { \gamma \over \gamma_{(0)} } \right )^{1/4}  =
2 \left( C + {E\over 3}   R_{(0)} \right ) \;
{\rm exp} \left (   \sqrt{3 \over 8} \phi \Bigg /
\left [ 1- { {8 E^2 \over 3} \overline R^{ab}_{(0)} \overline R_{ab}^{(0)}
\over
\left( C+ {E\over 3}  R_{(0)} \right )^2 }
\right]^{1/2}      \right ) \; ,
\end{equation}
\begin{equation}
z =   \phi \Bigg /
\left [ { 3 \left( C+ {E \over 3}  R_{(0)} \right )^2 \over
8 E^2 \overline R^{ab}_{(0)} \overline R_{ab}^{(0)} } - 1
\right]^{1/2}  \, ,
\end{equation}
\begin{equation}
[h]= [h^{(0)}] \; {\rm exp} \left [ - \sqrt {32  \over 3} E \,
{ [\overline R_{(0)}] \, [\gamma^{(0)}] \, \; \phi \over
\sqrt{ \left( C + {E \over 3}  \, R_{(0)} \, \right )^2
- { 8 E^2 \over 3 }  \overline R^{ab}_{(0)} \overline R_{ab}^{(0)} } }
\right ] \; .
\end{equation}
\endnumparts

Semiclassical evolution is very similar to the case with cosmological
constant, and it will not be discussed further.

\section{Conclusions}

In the semiclassical approximation, it was shown how to solve the system of
strongly-coupled gravity and matter using a Green's function
method.  There are two steps in implementing this program:

\noindent
{\it 1. Computing the Green's function.} Explicit Green's function solutions
of the energy constraint in a Hamilton-Jacobi context were given for systems
describing Einstein gravity interacting with either a dust field or a scalar
field.

\noindent{\it 2. Constructing the General Solution through the Superposition
Principle.}
One can construct a general semiclassical solution to the momentum constraint
as well as the energy constraint by a superposition over the parameter
fields of the Green's function.

The number of parameter fields appearing
in the Green's function is a tricky issue. Apparently, this number
should be one less than the number of fields originally
in the problem. For example, in the case of a dust field $\chi(x)$ interacting
with the 3-metric $\gamma_{ab}(x)$, there are
7 degrees of freedom per spatial point $x$: one for the dust field and
6 for the symmetric $3 \times3 $ matrix $\gamma_{ab}$. Hence, there should be
at most 6 parameter fields in the Green's function. If there are more,
one should set the additional parameter fields to zero.
Exact solutions demonstrate the validity of this approach.

The problem of strongly-coupled gravity interacting
with matter is not exceedingly complicated, although the actual details
may be intricate.
Many of the exact solutions presented in this paper had been derived earlier
using other techniques.
However, the Green's function method is very general
and in some sense all solutions to strongly-coupled gravity and matter
may be derived using it. The general methodology may perhaps
be useful in analyzing a wide range of gravitational phenomena.

\noindent
{\bf Acknowledgment}

\noindent
I would like to thank Don Page and Bill Unruh for making some useful 
suggestions.
This work was supported by the Natural Sciences and Research Council of
Canada (NSERC).

\References

\item[] Belinski V A, Khalatnikov I M and Lifshitz E M 1970
Oscillatory Approach to an Singular Point in the Relativistic Cosmology
{\it Advances in Physics} {\bf 19}, 525-73

\item[] Chiba T 1995 Applying Gradient Expansion to a Perfect Fluid and
Higher Dimensions, preprint KUNS-1277

\item[] Creutz M  1983 {\it Quarks, Gluons and Lattices}
(Cambridge University Press, Cambridge UK)

\item[] Darian B 1997 Solving the Hamilton-Jacobi equation for gravitationally
interacting electromagnetic and scalar fields {\it Class. Quantum Gravity}
(to be published)

\item[] Deruelle N and Langlois D 1995 Long Wavelength Iteration of Einstein's
Equations Near a Space-Time Singularity {\it Phys. Rev. D} {\bf 52}
2007-19

\item[] Dirac P A M 1958 The Theory of Gravitation in Hamiltonian Form
{\it Proceedings of the Royal Society} {\bf A246}, 333-43

\item[] Francisco G and Pilati M 1985 Strong-coupling quantum gravity. III.
Quasiclassical approximation {\it Phys. Rev. D} {\bf 31}, 241-50

\item[] Hartle J B 1997 {\it Quantum Cosmology: Problems for the 21st
Century}, in `Physics 2001', Nishinomiya-Yukawa Memorial Symposium on
Physics in the 21st Century: Celebrating the 60th Anniversary of the
Yukawa Meson Theory, Nishinomiya, Hyogo, Japan, 7-8 Nov 1996

\item[] Hartle J B and Hawking S W 1983 Wavefunction of the Universe
{\it Phys. Rev. D} {\bf 28}, 2960-74

\item[] Henneaux M, Pilati M and Teitelboim C 1982
Explicit Solution for the Zero Signature (Strong-Coupling) Limit
of the Propagation Amplitude in Quantum Gravity {\it Phys. Lett. B}
{\bf 110}, 123-28

\item[] Higgs P W 1958 Integration of Secondary Constraints in Quantized
General Relativity {\it Phys. Rev. Lett.} {\bf 1}, 373-74

\item[] Husain V 1988 The $G_{Newton} \rightarrow \infty$ Limit of Quantum
Gravity {\it Class. Quant. Grav.} {\bf 5}, 575

\item[] Landau L and Lifshitz E M 1960 {\it Mechanics}, p.148
(Pergamon Press, New York)

\item[] Landau L and Lifshitz E M 1975 {\it The Classical Theory of
Fields} (fourth English edition, Pergamon, Oxford)

\item[] Lifshitz E M and Khalatnikov I M 1964
{\it Usp. Fiz. Nauk} {\bf 80}, 391 [Sov. Phys. Usp.,  6, 495]

\item[] Moscardini L, Borgani S, Coles P, Lucchin F, Matarrese S,
Messina A and Plionis M 1993 Large Scale Angular Correlations
in CDM Models {\it Astrophys. J. Lett.} {\bf 413}, 55

\item[] Parry J, Salopek D S and Stewart J M, 1994
Solving the Hamilton-Jacobi Equation for General Relativity
{\it Phys. Rev. D} {\bf 49}, 2872-81

\item[] Peres A 1962 On the Cauchy Problem in General Relativity, II
{\it Nuovo Cim.} {\bf 26}, 53-62

\item[] Pilati M 1982 Strong Coupling Quantum Gravity. 1. Solutions in
a Particular Gauge {\it Phys. Rev. D} {\bf 26}, 2645

\item[] Polyakov A M 1987 {\it Gauge Fields and Strings}
(Harwood Academic Publishers, Switzerland)

\item[] Salopek D S 1991
Nonlinear Solutions of Long-Wavelength Gravitational Radiation
{\it Phys. Rev. D} {\bf 43}, 3214-33

\item[]\dash 1992a Cold-Dark-Matter Cosmology with Non-Gaussian Fluctuations
from Inflation {\it Phys. Rev. D} {\bf 45}, 1139-57

\item[]\dash 1992b Towards a Quantum Theory of Long-Wavelength Gravity
{\it Phys. Rev. D} {\bf 46}, 4373-86

\item[]\dash 1995 Characteristics of Cosmic Time
{\it Phys. Rev. D} {\bf 52}, 5563-75

\item[]\dash 1997 Coordinate-Free Solutions for Cosmological Superspace
{\it Phys. Rev. D.} {\bf 56}, 2057-64

\item[] Salopek D S and Bond J R 1990 Nonlinear Evolution of Long-Wavelength
Metric Fluctuations in Inflationary Models {\it Phys. Rev. D} {\bf 42}, 3936-62

\item[] Salopek D S and Stewart J M 1992 Hamilton-Jacobi Theory for
General Relativity with Matter Fields {\it Class. Quantum Grav.} {\bf 9},
1943-67

\item[]\dash 1993 Cosmological Fluids as Time Variables in General Relativity
{\it Phys. Rev. D} {\bf 47}, 3235-44

\item[] Salopek D S, Stewart J M and Parry J 1993
The Semi-Classical Wheeler-DeWitt Equation: Solutions for Long-Wavelength
Fields {\it Phys. Rev. D} {\bf 48}, 719-27

\item[] Soda J, Ishihara H and Iguchi O 1995 Hamilton-Jacobi Equation for
Brans-Dicke Theory and its Long Wavelength Solution
{\it Progress in  Theoretical Physics} {\bf 94}, 781-94

\item[] Teitelboim C 1982 Quantum Mechanics of the Gravitational Field
{\it Phys. Rev. D} {\bf 25}, 3159-79

\item[] Tomita K 1975 Evolution of Irregularities in a Chaotic Early Universe
{\it Progress in Theoretical Physics} {\bf 54}, 730

\item[] Veneziano G 1997 Inhomogeneous Pre-Big Bang String Cosmology,
CERN-TH/97-42, hep-th/9703150

\endrefs
\end{document}